\documentclass[11pt]{article}
\usepackage[super]{cite}

\usepackage{graphicx}
\usepackage{setspace}
\usepackage{fullpage}

\begin{document}

\title{``Narrow'' Graphene Nanoribbons Made Easier by
  Partial Hydrogenation}

  \author{
    Hongjun Xiang,$^{1*}$ Erjun Kan,$^{2}$ Su-Huai Wei,$^{1}$
    Myung-Hwan Whangbo,$^{2}$ Jinlong Yang$^{3}$  \\
    {$^1$ \it National Renewable Energy Laboratory, Golden, Colorado      80401, USA}\\    
    {$^2$\it Department of Chemistry, North Carolina State University}
    \\
    {\it Raleigh, North Carolina 27695-8204, USA} \\
    {$^3$ \it Hefei National Laboratory for Physical Sciences at  Microscale,}\\
    {\it University of Science and Technology of China, Hefei, Anhui 230026,    P. R. China}\\
{e-mail: hongjun\_xiang@nrel.gov} \\
}

\date{\today}

\maketitle

\begin{abstract}
  It is a challenge to synthesize graphene nanoribbons (GNRs) with narrow
  widths and smooth edges in large scale.
  Our first principles study on the  hydrogenation of GNRs
  shows that the hydrogenation starts from
  the edges of GNRs and proceeds gradually toward the middle of the
  GNRs so as to maximize the number of carbon-carbon $\pi$-$\pi$
  bonds. Furthermore, the partially hydrogenated wide GNRs have similar 
  electronic and magnetic properties as those of narrow GNRs.
  Therefore, it is not necessary to directly produce narrow GNRs for
  realistic applications because partial hydrogenation could make wide
  GNRs ``narrower''.  

\end{abstract}

\clearpage

\section*{Introduction}
Graphene, a two-dimensional (2D) single layer of carbon
atoms, is a rapidly rising star on the horizon of materials
science and condensed-matter physics.
It has attracted tremendous attention because of
its unique massless Dirac fermion-like electronic properties 
\cite{Neto2009} and potential applications
in electronic devices. \cite{Geim2007} 
When graphene is etched
or patterned along one specific direction, a novel
quasi one-dimensional (1D) structure, a strip of graphene of nanometers 
in width, can be obtained, which is referred to as
a graphene nanoribbon (GNR). The GNRs are predicted to
exhibit various remarkable properties and may be a potential
elementary structure for future carbon-based nanoelectronics. 
\cite{Son2006,Wakabayashi2001,Barone2006,Areshkin2007,Kan2008}
Remarkably, theoretical calculations \cite{Yang2007,Barone2006} predicted that quantum
confinement and edge effects make narrow
GNRs (width w $<$ 10 nm) into semiconductors,
which differs from single-walled carbon
nanotubes that contain some metallic
species. 
Thus, GNRs with narrow widths and atomically smooth edges could be
used as room temperature field
effect transistors with excellent
switching speed and high carrier mobility
(potentially, even ballistic transport).
Indeed, Li {\it et al.} \cite{Li2008} recently produced GNRs with width below 10 nm
using a chemical route and found that all of the sub-10-nm GNRs
were semiconductors and afforded graphene field effect transistors
with on-off ratios of about $10^7$ at room
temperature.\cite{Li2008,Wang2008} Unfortunately,
the yield of GNRs was low and their width distribution was broad;
widths ranged from less than 10 nm to $\sim$100 nm

To realize the practical potential of narrow GNRs, methods for
their mass production are sorely needed. 
Lithographic patterning has been used to produce wide ribbons ($>$20nm) from
graphene sheets, but the width and smoothness of the GNRs were
limited by the resolution of the lithography and etching techniques. \cite{Chen2007,Han2007}
Bulk amounts of wide (20$-$300nm) and few-layered GNRs were
synthesized by a chemical vapor deposition method.\cite{Delgado2008}
Very recently,
two research groups found ways to unroll carbon nanotubes to produce
nanoribbons. Jiao {\it et al.} \cite{Jiao2009} showed an approach to making GNRs
with smooth edges and a
narrow width distribution (10$-$20 nm) by unzipping
multiwalled carbon nanotubes (MWCNT) through plasma etching of nanotubes partly
embedded in a polymer film, but the ability of mass production through
this method is limited. 
On the other hand, 
Kosynkin {\it et al.} \cite{Kosynkin2009} described a simple solution-based oxidative process
for producing a nearly 100\% yield of water soluble nanoribbon structures by
lengthwise cutting and unravelling of MWCNT
side walls. The resulting ribbons are wider (around 100$-$500
nm), and not semiconducting, but easier to make  
in large amounts. Multilayered GNRs with a width about 100 nm were produced by
lithium intercalation and exfoliation of carbon nanotubes. \cite{Marquez2009}
The above experimental efforts indicate that a large scale production of
narrow GNRs is still very challenging.

In the present work, we propose that wide GNRs can be made 
semiconducting as in the case of narrow GNRs by partial hydrogenation, 
because the hydrogenation starts from
the edges of GNRs and proceeds gradually toward the middle of the
GNRs. In this way, 
the difficulty of directly synthesizing narrow and smooth GNRs can
be avoided. Besides the electronic device applications, certain partially hydrogenated GNRs with a 
band gap around 1.5 eV could be ideal materials for solar cell absorbers 
due to the high carrier mobility of GNRs.

\section*{Hydrogenation of 2D graphene}
As a first step, we examine the hydrogenation of 2D graphene. Full
hydrogenation of graphene has been extensively
studied,\cite{Sofo2007,Lu2009,Boukhvalov2008} but configuration of the partial 
hydrogenation is not well understood. It is well known that hydrogen
atoms tend to absorb on the top of carbon atoms.    
To find the lowest energy configuration of graphene with different
coverages of hydrogen atoms, we use the state of the art ``cluster
expansion'' method \cite{Ferreira1989} established in the alloy theory. 
In essence, the total energy of an alloy configuration is expressed by a
generalized Ising like model. The interaction parameters of the
cluster expansion are obtained by mapping the density functional total
energies of tens of different configurations to the generalized Ising
Hamiltonian. The Alloy Theoretic Automated Toolkit (ATAT) code \cite{atat} is 
employed here to construct the cluster expansion Hamiltonian.
The spin polarized density functional theory (DFT) calculations were
performed on the basis of the projector augmented wave method
\cite{PAW} encoded in the Vienna ab initio simulation package \cite{VASP} using the
local density approximation.
The internal coordinates and the cell of the sampling configurations
are fully relaxed  with  the plane-wave cutoff energy of 500 eV.

\begin{figure}
  \begin{center}
    \includegraphics[width=8.0cm]{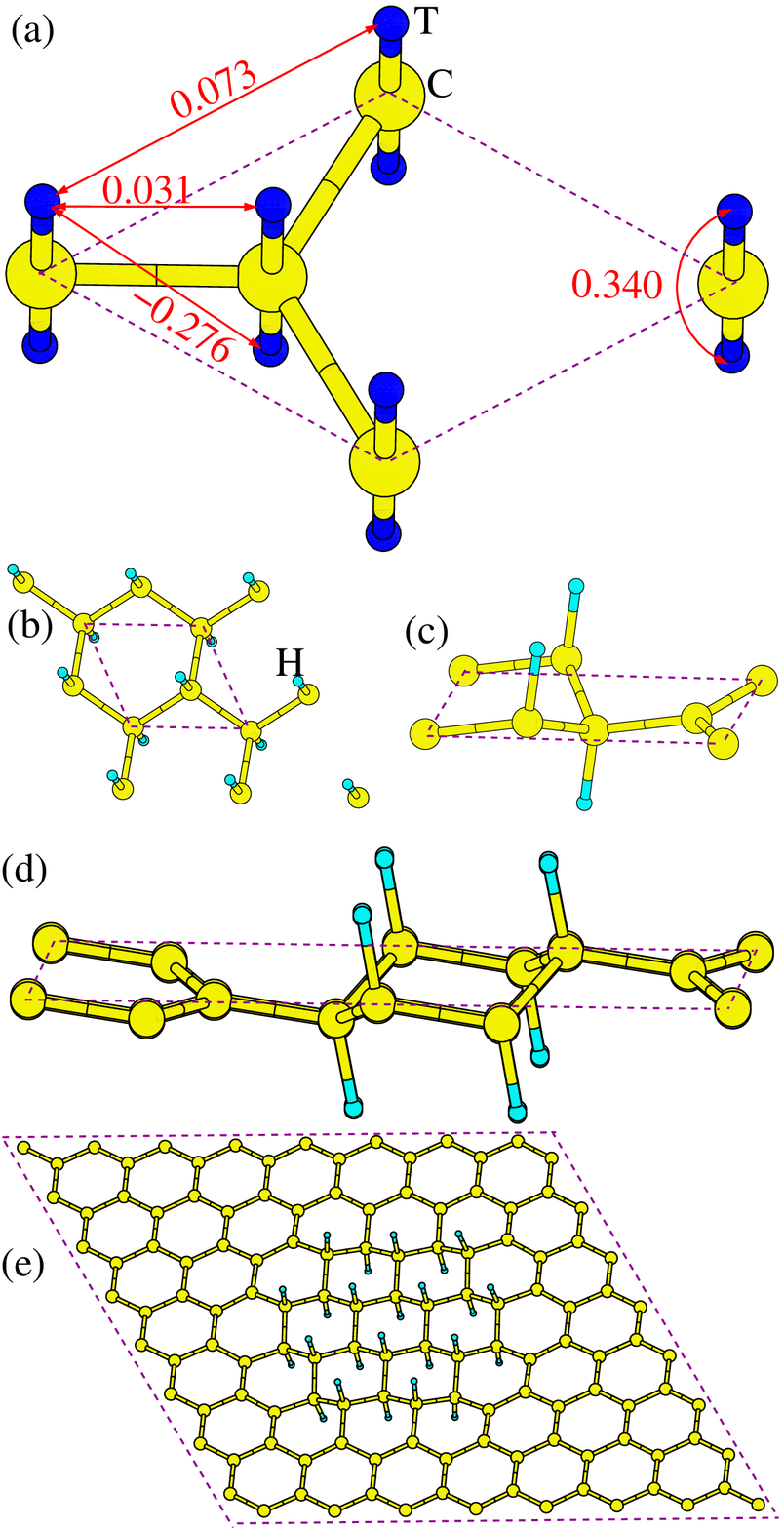}
    \caption{(a) The $C_2T_4$ structure used in the cluster expansion
    calculations. ``T'' refers to the top site of graphene.
    The important pair interactions are indicated by arrows. 
    The numbers (in eV) give the pair interaction parameters of the cluster
    expansion expression. (b) The ground state structure (graphane) of
    hydrogenated graphene with n[H]/n[C] $=1$.
    (c) The lowest energy structure of
    hydrogenated graphene with n[H]/n[C] $=0.5$ among all possible
    structures with no more than four carbon atoms per cell.
    (d) The lowest energy structure of
    hydrogenated graphene with n[H]/n[C] $=0.5$ among all possible
    structures with no more than eight carbon atoms per cell.
    (e) The ground state structure of a very large graphene cell with 24
    adsorbed H atoms. The enclosed region by dashed lines denotes  the cell. }
  \label{fig1}
  \end{center}
\end{figure}

In the cluster expansion process, we consider the C$_2$T$_4$ alloy
(see Fig.~\ref{fig1}a)
with the top sites T  occupied by H atoms or
vacancies. Our calculation shows that there are four important
pair interactions as shown in Fig.~\ref{fig1}a, and the three-body
interactions are negligible. We can see that the interaction between
two H atoms adsorbed on the same C atom is extremely repulsive
(0.340 eV). This is understandable because the five-fold coordinated
carbon is not stable. 
In contrast, we find that the interaction parameter
between two H atoms adsorbed on different sides of the two adjacent C
atoms is largely negative ($-0.276$ eV). The efficient strain
relaxation and the absence of dangling $\pi$ bonds in
this four-fold coordinated carbon 
configuration account for the stability. 
Using the cluster expansion Hamiltonian, we can easily obtain the
energy of a given alloy configuration and thus the ground state
structure of the partially hydrogenated graphene for a given
supercell. 
It is clear that the number of H atoms (n[H]) should not exceed that
of C atoms (n[C]) because otherwise some C atoms will bind with more than
one H atoms.
When n[H]/n[C] $= 1$, 
the ground state structure is graphane (see
Fig.~\ref{fig1}b), as found by Elias {\it et al.}. \cite{Elias2009} 
For n[H]/n[C] $= 0.5$, the lowest energy 
structure among all possible configurations with no more than 4 carbon 
atoms per supercell is shown in Fig.~\ref{fig1}c.
The adsorbed H atoms adsorb on a 1D zigzag carbon chain such that
each H atom has two neighbor H atoms adsorbed on the opposite side of
neighboring C atoms. This is consistent with the fact that the hydrogenation
of neighbor C atoms from opposite sides is  energetically  preferred.
However, we find that the above structure is not the global ground state structure 
with n[H]/n[C] $=0.5$. For example, the lowest energy
structure (Fig.~\ref{fig1}d) among all possible configurations with no
more than 8 carbon atoms per supercell has a lower energy by 44 meV/C
than does the structure shown in Fig.~\ref{fig1}c. 
This is so because the number of carbon-carbon $p_z$-$p_z$ $\pi$ bonds is
increased from 0.5/C (Fig.~\ref{fig1}c ) to 0.625/C (Fig.~\ref{fig1}d). 
It is expected that the global ground state
of the partially hydrogenated graphene displays macroscopic phase
separation between graphene and graphane regions.
To confirm this point, we consider a large supercell
($8\times 8$) with
24 adsorbed H atoms. Using the Monte Carlo annealing technique, we find all the H
atoms form a close packed cluster in the ground state configuration
(Fig.~\ref{fig1}e). 
This shows that, in the global ground state of a partially 
hydrogenated graphene, the phase separation into graphene and graphane
parts  would take place.

We also calculate the hopping barrier of 
an isolated H atom adsorbed on a carbon atom of graphene to 
the adjacent  carbon atom using the nudged elastic band method.\cite{Henkelman2000} The
calculated barrier is $0.84$ eV, which is close to the value ($\sim 0.7$ eV) in the
case of the (8,0) CNT.\cite{Zhang2007}
This suggests that isolated H atoms are relatively mobile, and  
thermal  annealing would result in the formation of 
the macroscopic H cluster.
Recently, Singh {\it et al.} \cite{Singh2009}
theoretically studied the electronic and magnetic properties of
``nanoroads'' of pristine graphene carved into the electrically insulating matrix
of fully hydrogenated carbon sheet. However, our results suggest that
it would be difficult to realize such patterned graphene nanoroads
because of the tendency for phase separation into graphene and graphane parts.

\section*{Hydrogenation of 1D graphene nanoribbons}
\begin{figure}
  \begin{center}
    \includegraphics[width=8.0cm]{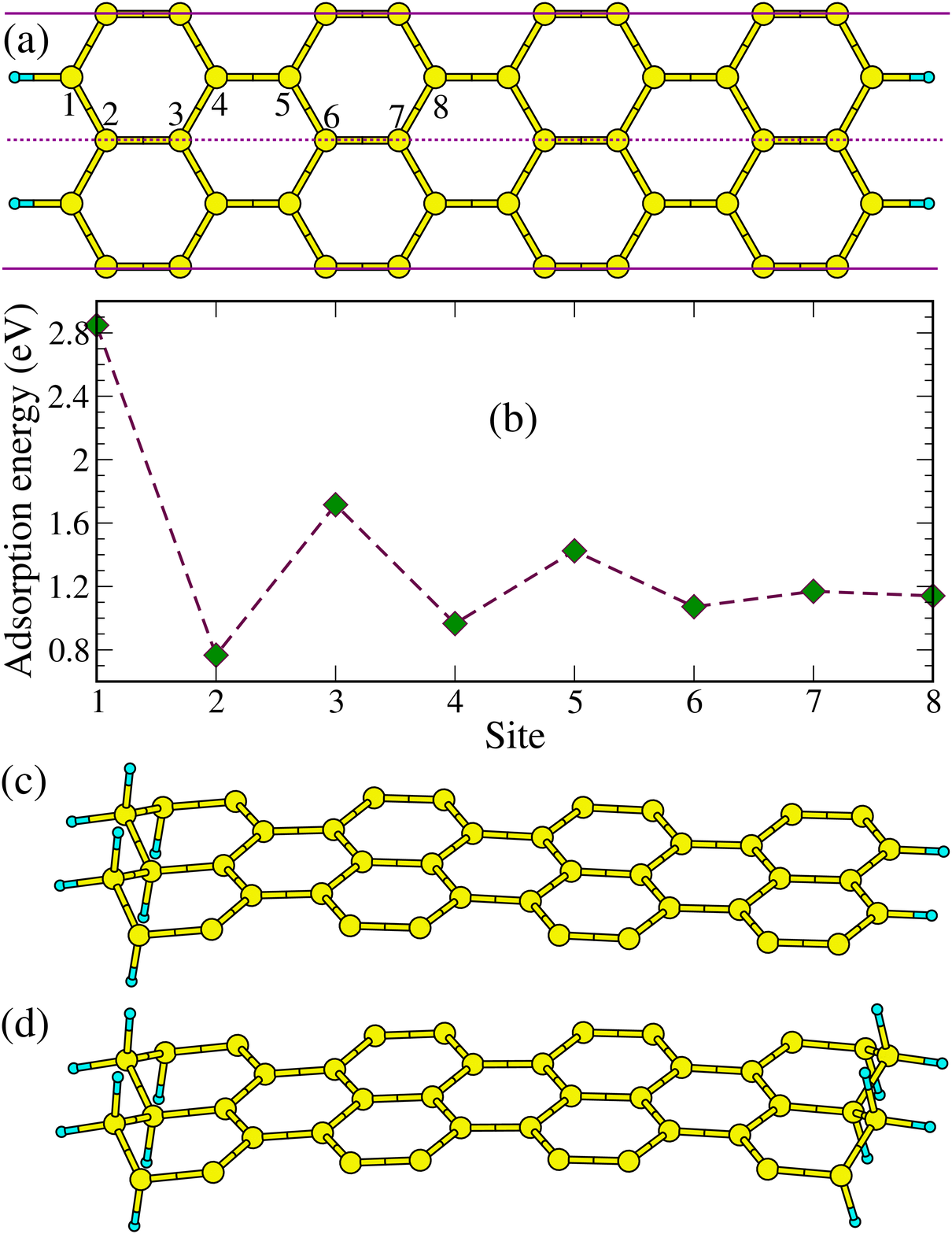}
  \caption{(a) The structure of 8-ZGNR with numbers indicating
    different nonequivalent carbon sites. 
    The supercell adopted in the DFT calculations is denoted by solid horizontal lines.
    (b) The adsorption energy of
    one H atom on different carbon sites as labeled in (a). 
    (c) and (d) show the ground state structures of 8-ZGNR with one
    and two adsorbed H atoms per edge carbon atom, respectively.}
  \label{fig2}
  \end{center}
\end{figure}

Now we turn to the study of hydrogenation of GNRs. 
There are two common types of GNRs. One kind of the GNRs, called
zigzag GNR (ZGNR), has zigzag-shaped edges with the
dangling $\sigma$ bonds passivated by hydrogen
atoms. Following conventional
notation, we name the GNR shown in
Fig.~\ref{fig2}a as 8-ZGNR according to the number of zigzag chains
along the ribbon. First, we consider the adsorption of a single H
atom on 8-ZGNR using the supercell with two unit cells. The adoption of a larger cell
leads to qualitatively similar results.
To find the most stable configuration, 
we consider all nonequivalent possible adsorption sites indicated in
Fig.~\ref{fig2}a. We define the adsorption energy as:
\begin{equation}
E_a=-[E\mbox{(H-GNR)}-E\mbox{(H atom)}-E\mbox{(GNR)}],
\end{equation}
where $E\mbox{(H atom)}$, $E\mbox{(GNR)}$, and $E\mbox{(H-GNR)}$ are
the total energies of an isolated H atom, the pristine GNR, and  the GNR with an
adsorbed H atom, respectively. Our calculated  adsorption energies are
shown in Fig.~\ref{fig2}b. The positive adsorption energy is a
consequence of the formation of C$-$H bond.
Remarkably, we find that the isolated H atom
prefers to adsorb on the edge carbon atom (site 1) than other sites by at least 1.1 eV. 
This is because the number of carbon-carbon $p_z$-$p_z$ $\pi$ bonds
is the largest in this configuration, similar to the
hydrogenated graphene case. 
Experimentally, it was found \cite{Wang2008B} that 
the atomic layer deposition of metal oxide on graphene grows actively
on edges, indicating that the chemical
reactivity at the edges of graphene is high,\cite{Wang2009} which is consistent with
our theoretical results.
Interestingly, the dependence of the
adsorption energy on the distance between the adsorbed site and the
edge is not monotonous: It displays an odd-even oscillation with a
smaller adsorption energy at even sites, and the adsorption energy of
even (odd) sites increases (decreases) with the distance, and
eventually the energy difference between even and odd sites adsorption
 becomes
very small. The smallest adsorption
energy  at site 2 might be due to the presence of two rather unstable edge carbon atoms
near site 2 that participate in the formation of only one $\pi-\pi$ bond. 

\begin{figure}
  \begin{center}
    \includegraphics[width=8.0cm]{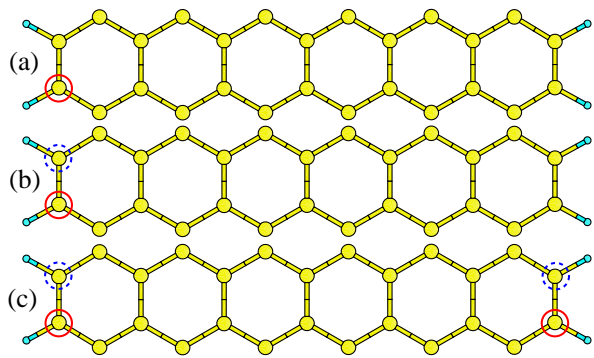}
  \caption{Schematic illustration of the ground state structures of 13-AGNR with (a) one or (b)
    two or (c) four adsorbed
    H atom per unit cell. The solid (dashed) big circle
    indicates that the H atom adsorbs on the top site above (below) the ribbon
    plane.}
  \label{fig3}
  \end{center}
\end{figure}

A second H atom will adsorb on the opposite
site of the carbon atom (site 2) adjacent to the edge carbon atom
(site 1) to which the
first H atom is bound, so as to saturate the broken bond. The
adsorption energy of this configuration is 5.73 eV/(two H atoms),
which is larger than the sum of the adsorption energies of a single H atom on site 1 and
2, indicating a cooperative adsorption behavior.  
This configuration is more stable than that with two H atoms
adsorbed on two outermost edge carbon atoms, for which the
adsorption energy (4.62 eV) is about twice the adsorption energy of a
single H atom on an edge carbon atom.
The third H atom is expected to adsorb on the edge carbon atom
adjacent to the carbon atom to which the
second H atom is bound. 
If the number of H atoms is equal to the total number of edge carbon
atoms (on both edges),
all the H atoms will
adsorb on the outermost carbon atoms of one edge
of the ZGNR, as shown in Fig.~\ref{fig2}c, where 
the left edge is assumed without loss of generality. 
When the number of H atoms is twice the total
number of edge carbon atoms, the excess H atoms will bind with 
the outermost carbon atoms of the right edge, resulting in a symmetric
configuration (see Fig.~\ref{fig3}d). The asymmetric configuration
where all H atoms adsorb on the left side has a higher energy by 30
meV per edge carbon atom. Nevertheless, the asymmetric configuration
has a similar electronic structure as the symmetric configuration
except for a small asymmetric splitting in the band structure.
Thus, our calculations suggest the following H
adsorption scenario: The H atoms adsorb on the outermost zigzag
bare carbon chain of one edge, and then on the outermost zigzag
bare carbon chain of the other edge. The process of alternating
hydrogenation continues until no more free H atom is available. 

In the above discussion, we focused on the hydrogenation of ZGNRs. 
To be complete, we now investigate the hydrogenation of another kind
of GNRs, namely, the armchair GNR (AGNR) with armchair-shaped edges. 
Similar to the case of ZGNRs, an isolated H atom also prefers to adsorb on
the edge carbon atom (see Fig. S1 for the calculated adsorption
energies), as shown in Fig.~\ref{fig3}a for the case of 13-AGNR 
case. A second H atom will adsorb on the adjacent carbon atom of
the edge C$-$C dimer,
as expected (Fig.~\ref{fig3}b). It is found that four H atoms also
adsorb symmetrically on the two edges of 13-AGNR (Fig.~\ref{fig3}c).  
Therefore the H adsorption on AGNRs resembles that on ZGNRs except that 
the H atoms adsorb on AGNRs in a dimer-line-by-dimer-line manner. 

\begin{figure}
  \begin{center}
    \includegraphics[width=8.0cm]{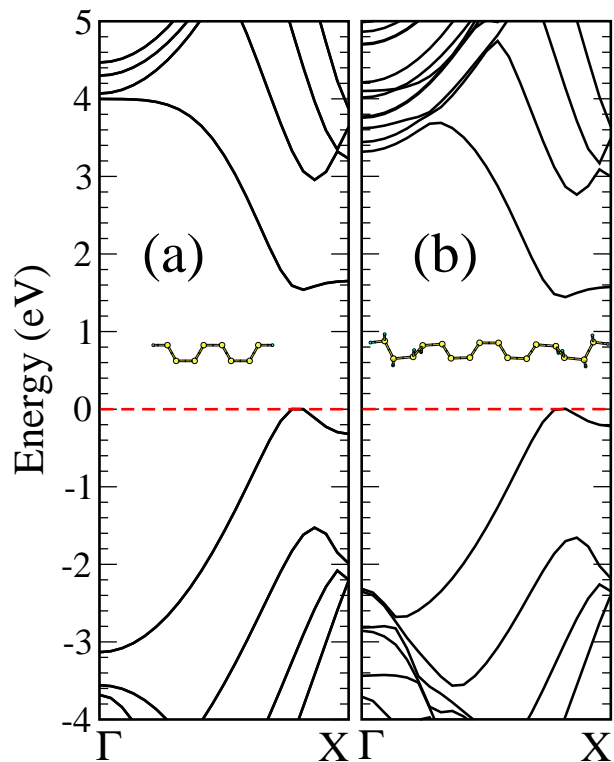}
  \caption{Electronic band structures of 4-ZGNR and 
    partially hydrogenated 8-ZGNR with four bare zigzag
    carbon rows (i.e., 8-ZGNR-4) in the AFM state from HSE06
    calculations. The horizontal dashed lines denote the top of the
    valence bands. The insets show the structures of 4-ZGNR and 8-ZGNR-4.}
  \label{fig4}
  \end{center}
\end{figure}

To understand the electronic and magnetic properties of
partially hydrogenated GNRs, we compare partially hydrogenated  8-ZGNR
with four bare zigzag carbon rows, hereafter referred to as 8-ZGNR-4,
with 4-ZGNR without adsorbed H atoms as a representative example.
In both cases, we find that the electronic ground state is the
antiferromagnetic (AFM) state in which each of the two electronic edge states 
is ferromagnetically ordered but the two edge states are 
antiferromagnetically coupled to
each other.  For 8-ZGNR-4, the AFM state is $-7.2$ meV/unit cell more stable than
 the ferromagnetic (FM) state, in which all spins are 
ferromagnetically aligned.
A similar stability difference between the AFM and FM
states is found for 
4-ZGNR (i.e., $-6.2$ meV/unit cell). 
Moreover, they almost have the same local magnetic moment ($\sim 0.10$
$\mu_B$). Bacause the local density
approximation  is well known to seriously underestimate the band gap of
semiconductors, we calculate the band structure of partially
hydrogenated ZGNRs by employing the screened
Heyd-Scuseria-Ernzerhof 06 (HSE06) hybrid functional,
\cite{Heyd2003,Krukau2006,Paier2006}  which was shown
to give a good band gap for many semiconductors including ZGNRs. \cite{Hod2008}
The HSE06 band structures calculated for 4-ZGNR and  8-ZGNR-4 in the AFM state are shown in
Figs.~\ref{fig4}a and b, respectively. In the energy region of the
band gap,  the band structure of 8-ZGNR-4 is similar to that of 4-ZGNR, and
the band gap (1.44 eV) of 8-ZGNR-4 is close
to that (1.53 eV) of 4-ZGNR. Therefore, the electronic and magnetic
properties of partially hydrogenated wide GNRs are determined by
those of its graphene part, i.e., the bare zigzag carbon rows. 
As already mentioned, 
semiconducting partially hydrogenated GNRs can be as transistors, and 
those with a small number of bare zigzag carbon rows
might be used as a solar cell absorption materials:  
8-ZGNR-4 (and N-ZGNR-4 with N$>4$) has a direct band gap that is 
close to the optimal value ($\sim 1.5$ eV) \cite{Thompson2008} for the
solar energy harvesting, and high carrier mobility.

\begin{figure}
  \begin{center}
    \includegraphics[width=8.0cm]{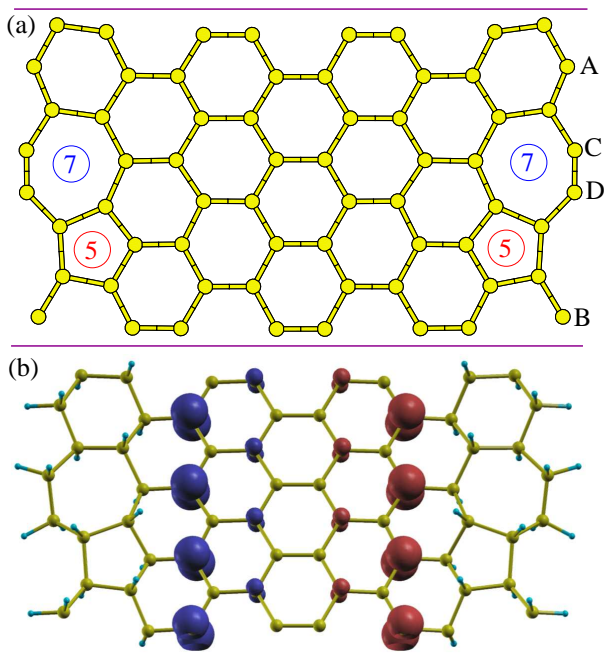}
    \caption{(a) Bare 8-ZGNR with Stone-Wales (SW) defects at the
      edges. 
      A, B, C, and D label different egde carbon atoms.
      The supercell is denoted by solid horizontal lines.
      (b) The partially hydrogenated 8-ZGNR containing
      SW defects with four zigzag bare carbon rows and the spin
      density distribution in the AFM state.}
  \label{fig5}
  \end{center}
\end{figure}

Experimentally, the edges of synthesized GNRs might be rough. 
It is interesting to see whether the resulting hydrogenated GNRs
with rough edges have a similar electronic structure as do hydrogenated
perfect GNRs. In order to address this issue, we study the
hydrogenation of bare 8-ZGNR with Stone-Wales (SW) reconstructions at the
edges  (see Fig.~\ref{fig5}a), which are typical defects for bare
GNRs. \cite{Huang2009} There are four edge carbon atoms per edge:
two (C and D in Fig.~\ref{fig5}a) of these belong to the 7-ring that
form a triplet bond with each other, and the others (A and B in
Fig.~\ref{fig5}a) are isolated edge carbon atoms of the 6-ring.  
Due to the presence of dangling $\sigma$ bonds at A and B sites, 
a single H atom will first bond to A or B site: The adsorption energy
calculation shows that $E_a\mbox{(A)}=7.40$ eV,
$E_a\mbox{(B)}=7.34$ eV, $E_a\mbox{(C)}=4.98$ eV, and $E_a\mbox{(D)}=5.01$ eV. 
Excess H atoms will adsorb gradually toward the inner part of GNRs.
Shown in Fig.~\ref{fig5}b is the partially hydrogenated 8-ZGNR containing
SW defects with four zigzag bare carbon rows. For the two carbon atoms
common to both 5-ring and 7-ring, two H atoms adsorb 
on them above the ribbon plane, as a result of the odd-membered
 ring. We find that this
partially hydrogenated 8-ZGNR has a similar properties as 8-ZGNR-4;
It is semiconducting with a similar band gap and the AFM state is more
stable than the FM state by $-5.9$ meV per ZGNR unit cell. The spin
density plot of the AFM state (Fig.~\ref{fig5}b) shows that the
magnetic moments are also mainly due to the sp$^2$ carbon atoms next
to the sp$^3$ carbon atoms. Therefore, partial hydrogenation can also 
convert a GNR with unsmooth edges into a GNR with perfect electronic
and magnetic properties.

\section*{Concluding remarks}
In summary, we performed a comprehensive first principles DFT study on
the hydrogenation of graphene and GNRs. The hydrogen adsorption on graphene
results in a complete phase separation between bare graphene and graphane. 
As for the hydrogen adsorption on GNRs, our study reveals the
following rules:
(i) Hydrogen atoms adsorb preferentially on the outermost edge carbon
atoms.
(ii) Hydrogen atoms lead to pairwise addition by adsorbing on adjacent
carbon atoms
in a one-up and one-down manner.
(iii) The above adsorption process shifts from one edge  to the other
edge of a GNR, and 
this alternating hydrogenation process continues until there is no
more free H atoms.

Our study suggests that the partial
hydrogenation can make wide GNRs effectively ``narrower'' in their 
physical properties, because
partially hydrogenated wide GNRs have electronic, optical, and
magnetic properties similar to those of the narrow GNRs representing 
their graphene parts.
Therefore, the experimental difficulty in synthesizing GNRs with narrow
width and smooth edges could be bypassed through partial
hydrogenation of wide GNRs. 
Partial hydrogenation might pave a way for the application of GNRs 
as transistors or novel carbon based solar cell absorption materials. 
In this study, we only consider the adsorption of H atoms. However,
the concept should remain valid if other groups are used
instead. 

\section*{Acknowledgments}
Work at NREL was supported by the U.S. Department of Energy, under
Contract No. DE-AC36-08GO28308, 
and work at NCSU by the U. S. Department of Energy, under Grant
DE-FG02-86ER45259.

\section*{Supporting Information Available:}
The adsorption energies of one H atom on different carbon sites of
13-AGNR. This information is available free of charge via the Internet
at http://pubs.acs.org/.

\clearpage

\clearpage
\ \\
\ \\
\ \\
\ \\
\ \\
\ \\

\begin{center}
  \includegraphics[height=5cm]{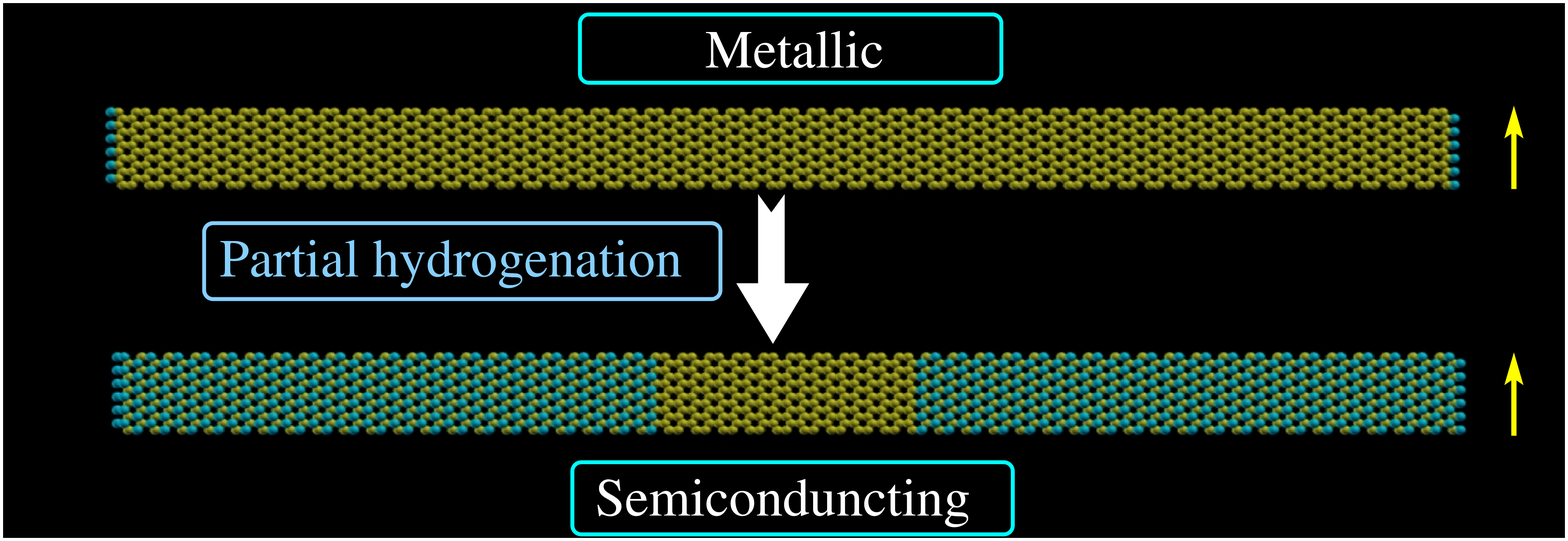}
\end{center}

\begin{center}
  TOC graphic
\end{center}

\end{document}